\documentclass[aip,jcp,preprint]{revtex4-1}
\usepackage{amsfonts}
\usepackage{amsmath,amssymb}
\usepackage{graphicx}
\usepackage{dcolumn}% Align table columns on decimal point
\usepackage{bm}% bold math
\draft % marks overfull lines with a black rule on the right
\usepackage{xcolor}

\DeclareMathOperator\erf{erf}

\draft % marks overfull lines with a black rule on the right

\begin{document}

\setlength{\parskip}{10pt}

\title{On the hydrodynamics of swimming enzymes} %Title of paper



\author{Xiaoyu Bai}
\email[]{Xiaoyu.Bai@rice.edu}
%\thanks{}
\affiliation{Department of Chemistry and Center for Theoretical Biological Physics, Rice university, Houston TX 77005 }

\author{Peter G. Wolynes}
\email[]{pwolynes@rice.edu}
%\thanks{}
\affiliation{Department of Chemistry and Center for Theoretical Biological Physics, Rice university, Houston TX 77005 }


\date{\today}

\begin{abstract}
Several recent experiments suggest that rather generally the diffusion of enzymes may be augmented through their activity. We demonstrate that such swimming motility can emerge from the interplay between the enzyme energy landscape and the hydrodynamic coupling of the enzyme to its environment. Swimming thus occurs during the transit time of a transient allosteric change. We estimate the velocity during the transition. The analysis of such a swimming motion suggests the final stroke size is limited by the hydrodynamic size of the enzyme. This limit is quite a bit smaller than the values that can be inferred from the recent experiments. We also show that one proposed explanation of the experiments based on reaction heat effects can be ruled out using an extended hydrodynamic analysis. These results lead us to propose an alternate explanation of the fluorescence correlation measurements.
\end{abstract}


\maketitle %\maketitle must follow title, authors, abstract and \pacs

\section{Introduction}

All swimming ultimately can be traced to the dynamics of enzymes. Both animal muscles and the cytoskeletons of individual  cells move through the cooperation of many motor proteins, which are enzymes that act together in large scale structures~\cite{howard2001, veigel2011}. Can individual enzymes, however, swim? How would they do so? The possibility of single particle molecular locomotion was already contemplated in setting up the theory of motorized crystals, motorized glasses and active molecular matter years ago~\cite{shen2004, wang2011jcp, wang2011pnas, wang2012pnas, wang2012jcp, wang2013jcp1, wang2013jcp2}. Nevertheless, we were surprised by recent observations~\cite{bustamante2015, mallouck2010} that suggest that a large range of enzymes, most of which are not in any way involved in biological motor activity, appear to swim, albeit in an undirected manner. Two groups have reported enhanced diffusion of several different enzymes that include catalase, urease and alkaline phosphatase. None of these are classical motor proteins. The apparent enhancement of diffusion appears to be proportional to enzymatic activity, just as is predicted by the theory of motorized assemblies~\cite{shen2004,wang2011pnas}. In their largely observational paper Riedel et al.~\cite{bustamante2015} also suggested a schematic mechanism by which the internal chemical energy in the substrate-enzyme complex could be transduced into motion of the enzyme's center of mass. They postulated the idea that the heat released by the reaction would lead to a pressure impulse in the surrounding water that in turn would lead to the motion of enzyme as a whole.

The reaction heat hypothesis was apparently inspired by the experimental observation that they found no enhanced diffusion for the enzyme triose phosphate isomerase, which catalyzes a reaction that does not release heat, while the other enzymes that catalyze reactions with larger $\Delta H's$ did apparently display activity enhanced diffusion. Although the quantitative details of the proposed locomotion mechanism were not completely laid out, on its face, the heat hypothesis itself raises some questions. For instance, if even momentarily the motion of the protein center of mass is supposed to be directed, why should the enthalpy change be the relevant thermodynamic quantity for determining the impulse, rather than a free-energy change? Also well established arguments suggest that the large scale motions of proteins should be highly damped by the solvent~\cite{wolynes1976, wolynes1977} and that therefore such motions of proteins should be described by the hydrodynamics of bodies at low Reynolds number and with low Mach number, the ratio of the characteristic speed of the moving object and the sound speed in the medium. The arguments put forward by Riedel et al. rely on the finite compressibility of the surrounding solvent and thus their picture entails high Mach number hydrodynamics.

In this paper we first explore an alternative explanation for how single enzymes might be able to swim while nevertheless only moving at low Reynolds and Mach numbers. The general problem of swimming at such low speeds has formed a long standing elegant part of biologically inspired physics starting with Purcell's so-called "Scallop Theorem"~\cite{purcell1977}. He used the theorem to point out the difficulties bacteria face in swimming. This theorem was later rigorously proved by Shapere and Wilczek~\cite{shapere1987}. The Scallop theorem states that to swim at low Reynolds number, the swim cycle must involve changing at least two degrees of freedom and also that the sequence of changes must not be time reversal invariant. The Scallop theorem constraints arise because the incompressible steady Navier-Stokes equations that describe fluid motion at low Reynolds number are time reversal invariant: reversing a forward stroke therefore causes the swimmer simply to re-trace its forward motion in the reverse sense so as to yield no net displacement~\cite{lauga2011, yeomans2014}. Nevertheless, a cyclic motion in two or more degrees of freedom need not be time reversal invariant so that asymmetric cyclic movements can allow a swimmer to crawl through its surroundings, no matter how viscous they are. Since enzymes are complicated molecules with many more than two degrees of freedom, and also recognizing that the enzymes degrees of freedom can be restored to equilibrium after release of a catalytic product without actually reversing the catalysis step itself, the Scallop Theorem does indeed allow an enzyme to translate forward in the process of carrying out a series of chemical reactions. Treating this problem for any specific enzyme in structural detail would doubtless be quite complex so here to make our conceptual point we confine ourselves to studying a very schematic model that envisions directly coupling the reaction coordinate of the reaction itself or a subsequent enzyme allosteric structural change coordinate describing motion after an enzymatic reaction to other degrees of freedom of the enzyme molecule that describe the relative motion of domains in the protein. While accounting explicitly for the dynamic stochastic coupling between reaction modes and the overall enzyme motion, this model is quite parallel to models of nanomoters already put forward that use deterministic cycles of swimmer shape change~\cite{najafi2004}. We see in this model that a key role is played by the hydrodynamics of coupling during the traversal period of the activated motion. The latter quantity has recently received much attention through experiments on protein folding~\cite{eaton2009, eaton2012}. Even the simplest form of the model, in our view, gives a plausible upper limit to the possible speed and stroke size of an enzyme moving through a solution no matter what the structural details of the enzyme cycle. In the most favorable imaginable case the maximum predicted stroke size from this estimate turns out to be smaller than the enzyme's hydrodynamic radius. We can easily imagine by involving more elaborate motional mechanisms that the final stroke size could actually be much smaller than the enzyme's hydrodynamic size, but as we see it, achieving stroke sizes significantly larger than the hydrodynamic radius would require a huge conformation change of the enzyme tantamount to its global folding. While actual biological motor motions are known to involve "cracking"~\cite{onuchic2007}, nothing as dramatic as complete unfolding has yet been contemplated, even for motor proteins, and indeed such unfolding seems even more unlikely for the particular enzymes that were studied experimentally, which are quite stable.
The upper bound character of our result therefore raises some difficulties for the present interpretation of the experiments: while a step size as big as the hydrodynamic radius is quite substantial and would clearly be adequate for most biological functional purposes, our predicted bound is much too small to explain the reported enhancements of diffusion; indeed even the maximum step size that our model would predict gives enhancements decidedly below the observability limits for the fluorescence techniques employed. The reported values of the diffusion enhancement for catalase in particular requires a step roughly 6 times larger than catalase's hydrodynamic radius. Clearly our theory cannot account for a step size with the large valued inferred from the experiments. As we shall show explicitly it is also hard to see how high Mach numbers can ever be achieved during enzyme locomotion so we also cannot see how the explanation based on the heating mechanism can be hydrodynamically plausible. Rather different mechanisms for enzyme locomotion have been proposed such as self-electrophoresis~\cite{paxton2006}. These mechanisms involve transiently modifying the composition of the solvent. Such effects are available for larger objects~\cite{lauga2011, yeomans2014, najafi2004} but these models, however, have subsequently been withdrawn as explanations for enzyme locomotion by their authors. At the end of this paper, we are therefore led to suggest an alternative interpretation of the fluorescence experiments whereby the measured changes in fluorescence do not in fact arise from enhanced diffusion due to enzyme swimming at all but rather arise from chemically specific sources of transient fluorescence quenching that have not been taken into account.

\section{The Swimming Enzyme Model}
\begin{figure}[H]
	\begin{center}
		\includegraphics[width=\textwidth]{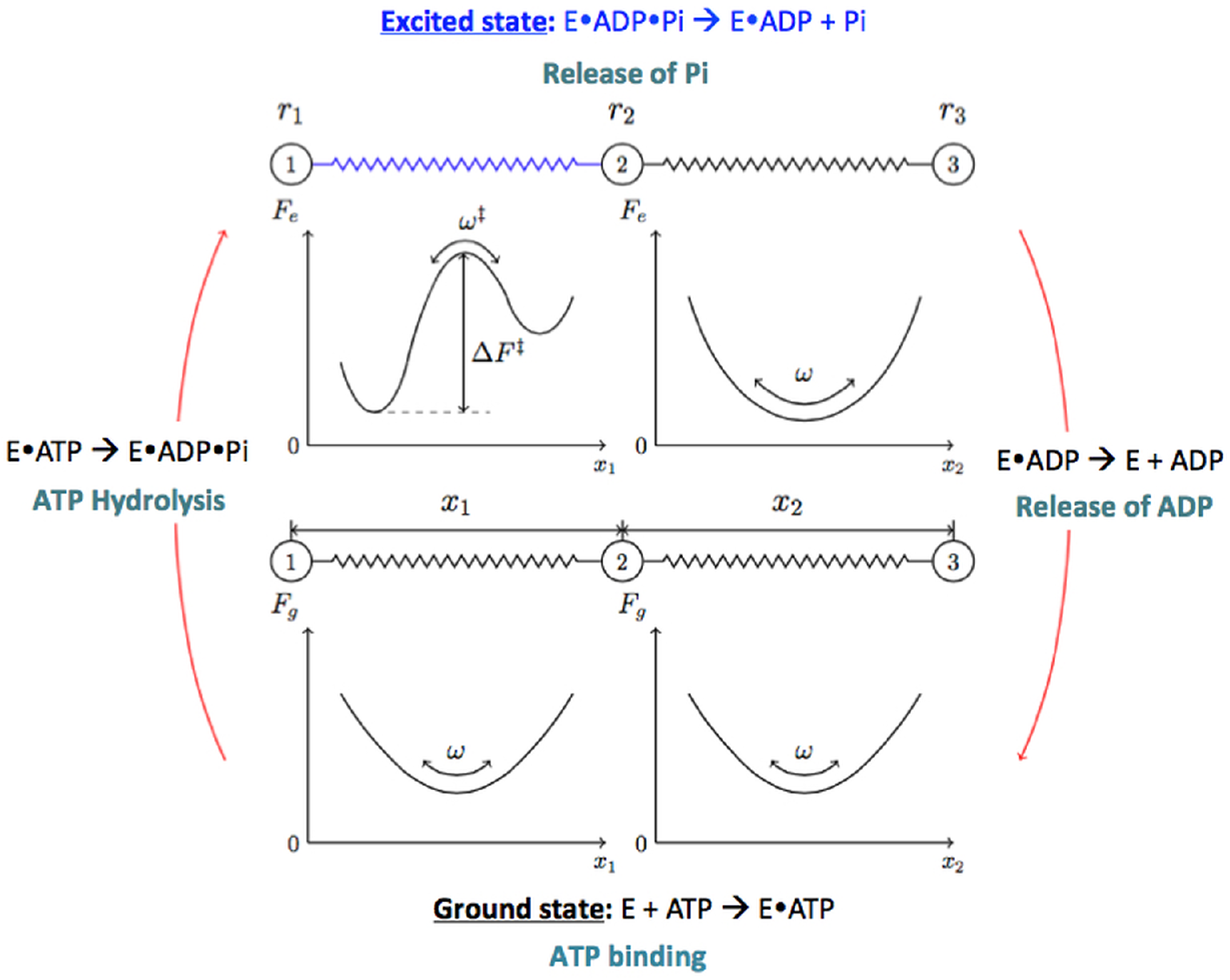} %depending on the latex compiler, you can omit the file extension
		\caption{Illustration of a 3-sphere model swimmer. $r_1$, $r_2$ and $r_3$ are coordinates for the three spheres in real 3D space. The panel above shows the free energy profile of excited state $F_e$ that projected onto internal coordinates $x_1$(left) and $x_2$(right), while the panel below shows the profile for the ground state $F_g$. All the curvatures are taken to be the same, $\omega = \sqrt{k/m}$ The instantaneous transition processes are denoted by the two red arrows. All the relevant chemical step are also indicated.}
		\label{fig:modelfig}
	\end{center}
\end{figure}

Our schematic model of a swimming enzyme envisions three protein domains that are hydrodynamically coupled to their surroundings. The relative motions of the domains are governed by two different free energy surfaces. One surface that describes the conformational changes accompanying ligand hydrolysis is called the "excited state" surface $F_e$ while the other, the so-called ground state surface $F_g$, describes the motion accompanying ligand binding and enzyme structural relaxation in the absence of hydrolysis. When these free energy surfaces are projected on the domain locations $r_1$, $r_2$ and $r_3$ we take the excited surface $F_e$ to be a bistable function of one of the interdomain distances, say $x_1$, as shown in Fig.\,\ref{fig:modelfig} while the relaxational ground state surface $F_g$ is taken to have a single minimum and is effectively harmonic. For simplicity we do not treat explicitly motion on the ground state surface and thus suppose the locomotion primarily occurs during the hydrolysis or allosteric step in the upper excited surface. Other possible schemes can be treated in a similar fashion to the present analysis. This set up is motivated by conformation switching models introduced earlier to describe allostery~\cite{itoh2010}. As pictured in Fig.\,\ref{fig:modelfig} the actual hydrolysis and release events are supposed to occur instantaneously so that the domains are considered to be fixed during the bond breaking events but the relevant motions of the domains after the chemical changes are generated by motion on either of the two surfaces  $F_g(x_1,x_2)$ and $F_e(x_1,x_2)$. The dynamics of the domain coordinates are coupled to the solvent hydrodynamically following the original scheme of Najafi and Golestanian~\cite{najafi2004}. Again for simplicity we take the domains to be equal size spheres with radius a that move one dimensionally along their mutual axis. The full cycle of motion therefore consists of stochastic instantaneous switching between these two surfaces interspersed with hydrodynamically coupled Brownian motion upon these two distinct surfaces. This Brownian motion involves the internal motion of the allosteric protein that ends up being frictionally coupled by hydrodynamics to the overall displacement of the enzyme.

The actual swimming motion occurs on the excited state surface by the internal coordinate of the enzyme stochastically leaping up from the initial configuration to the transition state configuration and then falling down to a product configuration. While making this leap fluid is displaced and the center of mass of the enzyme thereby moves. The most probable transition path can be found using by a path integral treatment that amounts to calculating the average of a path ensemble made up of all possible Brownian traversals over the free energy barrier. The details of this argument may be found in the Appendix. If we picture the bistable potential as being an inverted harmonic well with a spring constraint of $-k(k>0)$ the motions involved in traversing the free energy barrier can be found explicitly. Assuming equal and opposite negative spring constants for each domain as well as equilibrium spring length x for each domain the excited state barrier potential can be written as $F_e(x_1,x_2)=-\frac{1}{2}k(x_1 - x)^2+\frac{1}{2}k(x_2 - x)^2$. The hydrodynamically coupled Brownian Dynamics equations for the model swimmer in matrix form are:
\begin{equation}
	\label{eq:eom}
	\frac{d r_i}{dt} = \beta \sum_j D_{ij}F_j + \eta_i(t)
\end{equation}
where $F_i$ is the effective spring force exerted on each sphere and is determined by the excited potential $F_e$. $\eta_i(t)$ are white noise terms that are related to diffusion tensor $D_{ij}$ by the fluctuation-dissipation theorem $< \eta_i(t) \eta_j(t') >=2 D_{ij} \delta (t-t') $. Since the hydrodynamic coupling is kept at Oseen level as the first matrix inside the square bracket on the r.h.s. in equation \eqref{eq:eom}, the multiparticle diffusion tensor can be written as $D_{ij} = k_BT[\delta_{ij}\zeta_0^{-1} + (1- \delta_{ij} )T(r_{ij})]$, where $T(r) = (8\pi \eta_0 r)^{-1}( I + \frac{\hat{r} \hat{r}}{r^2} ) $ is the Oseen Tensor~\cite{wolynes1977diffusion}. To solve the coupled equations we carry out an eigen-analysis of the deterministic part for these linear equations. This is equivalent to extremizing the Onsager-Machlup Lagrangian so as to allow us to obtain the most probable path for a traversal from one well to the other. Since three degrees of freedom are involved  we find three eigenvalues. The eigenvalues for motion across the barrier consist of one zero eigenvalue $\lambda_1=0$ and two nonzero values: one stable $\lambda_2=-\sqrt{3} k(1-\frac{5a}{4x})/\zeta_0$ and the other unstable $\lambda_3=\sqrt{3} k(1-\frac{5a}{4x})/\zeta_0$. The existence of the zero frequency collective mode demonstrates the capability of net locomotion of the model swimmer. The dominant path for the collective mode (swimming motion) arises from this model and becomes simply:
\begin{equation}
	\label{eq:domp_1}
	r_1^\prime(t)=\frac{1}{3}[ r_1(t) + r_2(t) + r_3(t) ]=-\frac{7ka^2}{8\zeta_0x}t + x
\end{equation}
where the initial positions of the three spheres are set as  0, x and 2x respectively. We see the collective mode $r_1^\prime$ with the zero eigenvalue yields the motion of center of mass of the swimmer starting at location x as a uniform motion. Dominant paths for the auxiliary domains $r_2^\prime(t)$ and $r_3^\prime(t)$ are discussed in the SI. We also see that the swimming velocity during the transit time is $v_{swim}= -\frac{7ka^2}{8\zeta_0 x}$ clearly showing the hydrodynamic coupling of the swimmer to the solvent is crucial to its swimming behavior. How long does a given traversal take? This is essentially the time for the internal coordinate to go from the bottom of one well to the other. The transit time is obviously a distributed property. For the inverted oscillation model it is essentially the time for the reactive mode to move sufficiently to change the energy by an amount of $\Delta U^\ddagger$. The mean transit time therefore increases logarithmically with the increase of barrier height $\Delta U^\ddagger = \frac{1}{2} m \omega^\ddagger {x^\ddagger}^2$ since the internal coordinate shift on the inverted potential grows exponentially in time being described as the motion of an unstable damped system describing the domain motion. Typical traversal times have been discussed for the simplest models theoretically~\cite{hummer2004, zuckerman2007, makarov2010} and indeed have been measured experimentally for biomolecular folding process~\cite{eaton2009, eaton2012}. While the dominant path argument gives the most probable traversal time, one can also find explicitly the distribution for the transit time by normalizing the probability density flux from the pre-stroke to the post-stroke state with the flux being determined by the transit probability. This distribution can also be explicitly obtained from our path integral treatment. In the Appendix we provide the details of the derivation(see Appendix). The mean traversal time T determined from the distribution turns out to be a bit larger than the most probable transit time. Explicitly we find:
\begin{equation}
	\label{eq:transit_time}
	T \approx \frac{\ln(2\beta \Delta U^\dagger) + \gamma + \ln[\sqrt{3}(1 - \frac{5a}{4x}) /\cos^2{\phi}] }{\lambda_3}
\end{equation} 
where $\gamma$ is the Euler constant and $\phi$ is the angle between the reaction coordinate and the eigenvector associated with $\lambda_3$ in the euclidean space spanned by $r_1$, $r_2$ and $r_3$. This angle follows from the hydrodynamic coupling as described in the Appendix. Knowledge of the free energy barrier and this eigenvalue corresponding to the unstable mode is enough for evaluating the mean transit time T, but determining the coupled dominant leaping path demands the motions in the pre-stroke and post-stroke state of enzymes, which we neglect here. A simple direct prediction for the stroke size L possible at the Oseen limit $(\frac{a}{x}\rightarrow 0)$ follows from this estimation for the mean transit time combined with the velocity of the collective mode that results from the path-integral treatment:
\begin{align}
	\label{eq:strokesize}
	 L=|v_{swim} \times T| & = \frac{7a^2[\ln{(2 \beta \Delta U^\ddagger)} + \gamma + \ln(\sqrt{3}/\cos^2{\phi})]}{8\sqrt{3}x} 
\end{align} 
The Euler constant $\gamma$ is small and the free energy barrier for a typical enzymatic reaction is just several $k_BT$. For most enzymes, the size of each subunit $a$ is usually comparable to the separation distance $x$ and thus, one can infer from the equation \eqref{eq:strokesize} that the size of the net displacement can only be of the same order as the hydrodynamic size of enzymes, e.g. $L \sim \frac{a^2}{x} \sim a$. From this formula one can estimate the maximum stroke sizes. We do this for catalase, one of the four enzymes upon which the fluorescence correlation measurement have been conducted by Riedel et al~\cite{bustamante2015}. The hydrodynamic radius of catalase is 5.22nm. The stroke sizes for catalase assuming three different Oseen ratios $\frac{a}{x}$ = 0.1, 0.5 and 1 turn out to be correspondingly $9.2\times10^{-2}$nm, 2.3nm and 9.2nm where $\Delta U^\ddagger$ is taken as $5k_BT$. Riedel et al. did not report stroke size, but the theory of motorized assemblies suggests that the slope $\alpha$ of the plot of Diffusion constant vs. the enzyme turnover rate is essentially the stroke size~\cite{wang2011jcp}. The stroke size for catalase determined from the measured values of $\sqrt{\alpha}$ is as large as 31.6nm. This discrepancy between the hydrodynamic model and the experiments~\cite{bustamante2015}, argues that a careful examination of other possible channels of fluorescence correlation decay besides enhanced enzyme diffusion must be considered. 

We wish now to show explicitly that due to damping the motions of a swimming enzyme cannot excite a significant acoustic response in the fluids  as was envisioned in the reaction heat mechanism proposed by Riedel et al. To treat the hydrodynamic effects beyond the low Mach number regime, one only needs to replace the long-time limit of the friction $\zeta_0$ (given in our previous analysis using only steady hydrodynamics) with its frequency-dependent counterpart $\zeta(\omega)$. The frequency-dependent friction coefficient accounts both for finite momentum diffusivity and for sound propagation. Wolynes and McCammon~\cite{wolynes1977} calculated $\zeta(\omega)$ for a biopolymer decades ago by using nonsteady hydrodynamics. To focus on the acoustic effects alone one simply needs to omit the viscous contribution from their expression for $\zeta(\omega)$. This simplification gives a pure acoustic drag coefficient at zero-frequency with the value $\zeta(0)=\zeta_0\frac{a c_0}{9\nu_0}$. When substituted again into the long-time limit equation of motion to obtain the swimming velocity we find the pure acoustic mode would be even more strongly coupled to the motion of the swimmer than the viscous modes are. Thus if sound plays a significant role, the enzyme motion would be actually more strongly damped than it was our calculation where only the viscous coupling is taken into account. The acoustic effects clearly reduce the swimming speed: the swimming velocity when the pure acoustic effect is dominant turns out to be $v_{sound}=\frac{9\nu_0}{a c_0}v_{swim}$. That the acoustic effect is small and that it does not influence the locomotion is not unexpected. The perturbed pressure field usually relaxes more rapidly to its steady value than the perturbed velocity field such that the prefactor of $v_{sound}$ as $\frac{\nu_0}{a c_0} =\frac{a/c_0}{a^2/\nu_0}=\frac{\tau_s}{\tau_\nu}$ is also small, where ${\tau_s}$ and ${\tau_\nu}$ are the typical time for sound wave to propagate a distance of $a$ and momentum to diffuse over an area of $a^2$ respectively. That ${\tau_s}$ is much smaller than ${\tau_\nu}$ supports our prediction that compression indeed should be a small correction to the locomotion speed.

\section{Conclusion}

While we see that individual enzymes can swim randomly when they carry out a catalytic reaction and that therefore chemistry can in principle lead to enhanced diffusion of an individual enzyme proportional to the enzyme's activity, the effect should be quite a bit smaller than the recent experiments seem to suggest. We therefore think alternate explanations of the data need to be entertained. The absolute measured changes of the fluorescence correlations are quite subtle, so other ways of losing fluorescence correlation other than through enhanced diffusion out of the illuminated region should be considered. One possible way to lose fluorescence correlation is for an intermediate in the catalysis or a reaction product of the enzymatic reaction to quench the fluorescence when an active species forms transiently near the monitored fluorophore. In that case an extra source of decorrelation would be present and it would be proportional to the enzyme activity just as the proposed diffusion enchanced effect is. It seems essential to rule out this possibility for the alkaline phosphatase system which by catalyzing the hydrolysis of nitrophenylphosphate yields a fluorescently active product nitrophenol. Likewise the two enzymes urease and catalase have spectroscopically and electronically active metal centers. As these centers undergo the catalytic cycle they might yield species that can transiently quench the fluorescence, most likely through electron transfer. In order to estimate the size of these quenching effects much detailed spectroscopic electrochemical and kinetic data would be needed. This is beyond the scope of our present effort. It is most interesting, however, that the one system that displayed no excess decorrelation, triose phosphate isomerase, also lacks metal centers and catalyzes a reaction that would involve no active products. For this system then the activity induced quench mechanism could not operate and thus would explain why no enhanced diffusion was observed in that case. We point out that the excess transient quenching mechanism can be experimentally distinguished from the enhanced diffusion mechanism by carefully measuring the effect of changing the size of the illuminated region. Such a change should alter the diffusion signal but not the transient quenching signal.

As we finished preparing this account for publication, a theoretical study by Golestanian~\cite{golestanian2015} appeared that supports the idea that stochastic swimming can contribute to the enzyme activity enhancement of diffusion, but like us he concludes that the effects should be much smaller than those that were measured. In seeking an explanation of the observation he points out that the global heating by the enzyme reaction could result in heating up the whole sample which would appear as enhanced diffusion. The original authors of the experiments had considered but ruled out such an effect in their work. Obviously besides checking out the possibility of excess transient quenching we hope that future experiments with accurate temperature monitoring will also be undertaken. We look forward to such experimental investigations.

\begin{acknowledgements}
We thank Jose Onuchic for a most helpful discussion. This work was supported by the Center for Biologically Physics sponsored by National Science Foundation Grants PHY-1308264 and PHY-1427654. Additional support was provided by the D. R. Bullard-Welch Chair at Rice University Grant C-0016.
\end{acknowledgements}

\appendix

\section{Path-integral Treatment for Brownian Crossing of a Inverted Harmonic Barrier}
Brownian dynamics within the inverted harmonic potential $U(x) = -\frac{1}{2} m\omega^{\ddagger^2} x^2 + \Delta U^\ddagger$ is described by an overdamped Langevin equation:
\begin{equation}
	\frac{dx}{dt} = b x + \eta(t),\quad <\eta(t)\eta(t')> = 2D\delta (t-t')
\end{equation}
where $b = \beta D m \omega^{\ddagger^2} $ is the frequency involved in overdamped motion. To describe the probability functional for a Brownian path one has the Onsager-Marchlup Lagragian associated with the Brownian motion in the inverted  harmonic potential~\cite{wio2013}: 
\begin{equation}
	\mathcal{L}(\dot{x}, x, t) = \frac{1}{4D} (\dot{x} - bx)^2 + \frac{b}{2}
\end{equation}
The corresponding Euler-Lagrange equation for the dominant path $\frac{d}{dt} \frac{\partial \mathcal{L} }{\partial \dot{x}} - \frac{\mathcal{L}}{\partial x} = 0$ then is:
\begin{equation}
	\label{eq:eleqn}
	\ddot{x} - b^2 x = 0
\end{equation}
This equation of motion for x describes an unstable damped motion. Given the initial conditions $(t=0)$ and the final positions $(t=T)$ of the diffusing degree of freedom x as $x(0)$ and $x(T)$ respectively, one finds the most probable path for x(t):
\begin{equation}
	\label{eq:domipath}
	x(t) = \frac{x(T) \sinh (bt) + x(0) \sinh [b(T-t)]}{\sinh (bT)}
\end{equation}
The most probable leaping path starts from the initial position $x(0) = -\sqrt{ \frac{2 \Delta U^\ddagger}{m \omega^{\ddagger^2} } }$ proceeding to the final position $x(T) = +\sqrt{ \frac{2 \Delta U^\ddagger}{m \omega^{\ddagger^2} } }$. At these endpoints the free energy has fallen by $\Delta U^\ddagger$ from the barrier top. Thus the dominant leaping path for traversing in a time $T$ is given by:
\begin{equation}
x(t) = \sqrt{ \frac{2 \Delta U^\ddagger}{m \omega^{\ddagger^2} } } \frac{ \sinh (bt) - \sinh [b(T-t)] }{\sinh (bT)} 
\end{equation}

\section{Transit time in 1D inverted harmonic potential}

From the dominant path equation~\eqref{eq:domipath}, one can calculate the conditional transition probability density $\Phi(x, t|x_0, 0)$. This is also the propagator of a 1D Fokker-Planck equation with absorbing boundary conditions at the start and end of the trajectory at the bottom of the inverted well:
\begin{align}
	\Phi(x, t|x_0, 0) & = \int \mathcal{D}[x(t)] \exp{ [ -\int_0^t d\tau \mathcal{L}(\dot{x}, x, \tau) ]} \\
	& = [\frac{b}{2\pi D (e^{2bt} - 1)}]^{\frac{1}{2}} \exp{ [- \frac{b(x - x_0 e^{bt})^2}{2D(e^{2bt} - 1)}] }
\end{align} 
Recalling that the Fokker-Planck equation is a probability balance equation we see the probability flux for starting at $x_0$ at $t=0$ while ending up at $x$ at time t is proportional to the probability of the transit time being t:
\begin{align}
	\frac{\partial \Phi(x, t|x_0, 0)}{\partial t} & = - \frac{\partial}{\partial x} J(x, t|x_0, 0)\\
	J(x, t|x_0, 0) & = - D e^{\beta U(x)} \frac{\partial}{\partial x}[ e^{-\beta U(x)} \Phi(x, t|x_0, 0)]
\end{align}
We will denote the initial and final position $\pm \sqrt{\frac{2\Delta U^\ddagger}{m \omega^{\ddagger^2}}}$ as $\pm \Delta x$ for convenience. Thus the reactive flux $J(\Delta x, T| -\Delta x, 0)$ with transit time $T$ is:
\begin{equation}
J(\Delta x, T| -\Delta x, 0) = \frac{b \Delta x}{4} \sqrt{ \frac{b}{\pi D} } \frac{\exp [-\frac{b \Delta x^2}{2D} \coth(bt/2)]}{\sinh(bt/2) \sqrt{\sinh(bt)}}
\end{equation} 
The probability distribution density $P(t)$ for transit time is just the normalized reactive flux:
\begin{align}
	P(t) & =\frac{J(\Delta x, t|-\Delta x,0)}{\int_0^\infty J(\Delta x,t|-\Delta x,0) dt}\\
	& = \frac{b \Delta x}{2[1 - \erf(\Delta x \sqrt{b/(2D)})]} \sqrt{\frac{b}{\pi D}} \frac{\exp [-\frac{b \Delta x^2}{2D} \coth(bt/2)]}{\sinh(bt/2) \sqrt{\sinh(bt)}} \\
	& = \frac{b\sqrt{\beta \Delta U^\ddagger} }{1-\erf(\sqrt{\beta \Delta U^\ddagger})} \frac{\exp [-\frac{b \Delta x^2}{2D} \coth(bt/2)]}{\sinh(bt/2) \sqrt{\sinh(bt)}}
\end{align}

The mean transit time $T$ can now be obtained as the first moment of $P(t)$ in the long-time limit $(bt>>1)$:

\begin{align*}
 T & =\int_0^\infty t P(\tau)dt\\[10pt]
 & \approx  \int_0^\infty \tau d\tau \frac{b\sqrt{\beta\Delta U^\ddagger}}{e^{-\beta\Delta U^\ddagger}/\sqrt{\pi\beta\Delta U^\ddagger}}\times
   \frac{\exp{[-\beta\Delta U^\ddagger (1+2e^{-bt})]} }{e^{bt/2}/2\times \sqrt{\pi e^{bt}}}\\[10pt]
 & =\int_0^\infty \tau d\tau \times2b\beta\Delta U^\ddagger \exp{[-b\tau-2\beta\Delta U^\ddagger e^{-b\tau}]}\\[10pt]
 & \approx \frac{1}{b} \ln{[(2\beta\Delta U^\ddagger)+\gamma]}\quad (when\ \beta\Delta U^\ddagger >>1)
\end{align*}

 where two approximations have been made: i), the free-energy barrier is high such that $\beta \Delta U^\ddagger >>1$; ii), $coth(x) =1 + \frac{2 e^{-x}}{1 - e^{-x}} \approx 1 + 2 e^{-x}$ when $x$ is large. From the distribution function, one can also calculate that the most probable transit time $t_{m.p.}$, which differs from mean transit time $T$ by the term containing the Euler constant: $t_{m.p.} = \frac{1}{b} \ln(2\beta \Delta U^\ddagger)$.

\section{The most probable traversal path for a composite system: the model swimming enzyme}
We start with the equation of motion for the coupled Brownian dynamics for the three domains of the enzyme:
\begin{equation}
\frac{dr_i}{dt} = \beta \sum_j D_{ij}F_j + \eta_i(t)
\end{equation}
$F_i$ is the effective spring force exerted on each sphere and is determined by the excited potential $F_e$. $\eta_i(t)$ are white noise terms that are related to the multiparticle diffusion tensor $D_{ij}$ by the fluctuation-dissipation theorem $< \eta_i(t) \eta_j(t') >=2 D_{ij} \delta (t-t') $. At the Oseen level the diffusion tensor $D_{ij}$ is $D_{ij} = k_BT[\delta_{ij}\zeta_0^{-1} + (1- \delta_{ij} )T(r_{ij})]$, where $T(r) = (8\pi \eta_0 r)^{-1}( I + \frac{\hat{r} \hat{r}}{r^2} ) $ is the Oseen Tensor. This set of equations can be expressed more explicitly as:
\begin{align}
\frac{d}{dt}\begin{pmatrix} r_1 \\ r_2 \\ r_3 \end{pmatrix} = & \frac{1}{\zeta_0}
	\begin{pmatrix}
	1 & \frac{3a}{2x} & \frac{3a}{4x} \\
	\frac{3a}{2x} & 1 & \frac{3a}{2x} \\
	\frac{3a}{4x} & \frac{3a}{2x} & 1
	\end{pmatrix} \Big[
	\begin{pmatrix}
	k & -k & 0 \\
	-k & 0 & k \\
	0 & k & -k
	\end{pmatrix} 
	\begin{pmatrix} r_1 \\ r_2 \\ r_3 \end{pmatrix} + \begin{pmatrix} kx \\ -2kx \\ kx \end{pmatrix} \Big] +
	\begin{pmatrix} \eta_1(t) \\ \eta_2(t) \\ \eta_3(t) \end{pmatrix} \\[10pt]
	\label{eq:eommatrix}
	= & \frac{k}{\zeta_0}
	\begin{pmatrix}
	1 - \frac{3a}{2x} & -1 + \frac{3a}{4x} & \frac{3a}{4x} \\
	-1 + \frac{3a}{2x} & 0 & 1 - \frac{3a}{2x} \\
	-\frac{3a}{4x} & 1 - \frac{3a}{4x} & -1 +\frac{3a}{2x}
	\end{pmatrix}
	\begin{pmatrix} r_1 \\ r_2 \\ r_3 \end{pmatrix}
	+ \frac{k}{\zeta_0} \begin{pmatrix} x - \frac{9}{4}a \\ -2x + 3a \\ x - \frac{9}{4}a \end{pmatrix} 
	+ \begin{pmatrix} \eta_1(t) \\ \eta_2(t) \\ \eta_3(t) \end{pmatrix}\\
	= & \textbf{A\ r} + \textbf{b} + \boldmath{\eta(t)}
\end{align}

Diagonalizing the composite force-mobility matrix part leads us to three real eigenvalues associated with three eigen-modes $r_1^\prime$, $r_2^\prime$ and $r_3^\prime$. These eigenvalues are also the three eigen-frequencies of the swimmer's dynamics. By replacing the spring constraint b in the 1D traversal time path with the eigenvalues, the dominant path for these independent eigen-modes $r_1^\prime$, $r_2^\prime$ and $r_3^\prime$ can be found as well as the paths for the original coordinates $r_1$, $r_2$ and $r_3$. The resulting eigenvalues and eigenvectors are summarized as follows:
\begin{align*}
  \textbf{Eigenvalues: } & \lambda_1=0,\quad \lambda_2=-\sqrt{3} \frac{k}{\zeta_0} (1-\frac{5a}{4x}),\quad \lambda_3=\sqrt{3}\frac{k}{\zeta_0}(1-\frac{5a}{4x})\\[10pt]
  \textbf{Eigenvectors: } & \begin{pmatrix} 1 \\ 1 \\1 \end{pmatrix},\quad
  \begin{pmatrix}
  -2+\sqrt{3}+\frac{a}{x}(3-2\sqrt{3})\\
  1-\sqrt{3}+\frac{a}{x}(\frac{-9+5\sqrt{3}}{4})\\
  1 
  \end{pmatrix},\quad
  \begin{pmatrix}
  -2-\sqrt{3}+\frac{a}{x}(3+2\sqrt{3})\\
  1+\sqrt{3}+\frac{a}{x}(\frac{-9-5\sqrt{3}}{4})\\
  1 
  \end{pmatrix}
\end{align*}

The eigenmodes $r_i^\prime(t)$ can be transformed to the original coordinates $r_i(t)$ through a transformation matrix \textbf{P} composed of the eigenvectors:
\begin{align}
	\begin{pmatrix} r_1^\prime \\ r_2^\prime \\ r_3^\prime \end{pmatrix} 
	& = \textbf{P}^{-1} \begin{pmatrix} r_1 \\ r_2 \\ r_3 \end{pmatrix}
	=  \begin{pmatrix}
		1 & -2+\sqrt{3}+\frac{a}{x}(3-2\sqrt{3}) & -2-\sqrt{3}+\frac{a}{x}(3+2\sqrt{3}) \\
		1 & 1-\sqrt{3}+\frac{a}{x}(\frac{-9+5\sqrt{3}}{4}) & 1+\sqrt{3}+\frac{a}{x}(\frac{-9-5\sqrt{3}}{4})\\
		1 & 1 & 1
	\end{pmatrix}^{-1} \begin{pmatrix} r_1 \\ r_2 \\ r_3 \end{pmatrix}\\[10pt]
	& =   \begin{pmatrix}
	\frac{1}{3}+\frac{a}{12x} & \frac{1}{3}-\frac{a}{6x} & \frac{1}{3}+\frac{a}{12x} \\
	-\frac{1}{6}-\frac{(1-3\sqrt{3})a}{24x} & -\frac{1+\sqrt{3}}{6}+\frac{(1-\sqrt{3})a}{12x} & \frac{2+\sqrt{3}}{6}-\frac{(1+\sqrt{3}) a}{24x} \\
	-\frac{1}{6}-\frac{(1+3\sqrt{3})a}{24x} & -\frac{1-\sqrt{3}}{6}+\frac{(1+\sqrt{3})a}{12x} & \frac{2-\sqrt{3}}{6}-\frac{(1-\sqrt{3}) a}{24x}
	\end{pmatrix} \begin{pmatrix} r_1 \\ r_2 \\ r_3 \end{pmatrix}\	
\end{align}

The collective mode $r_1^\prime$ with zero eigenvalue is apparently the motion for center of mass at the Oseen limit $(\frac{a}{x}\rightarrow0)$: $r_1^\prime = \frac{1}{3}(r_1 + r_2 + r_3)$. Transformation of the constant part of equation~\eqref{eq:eommatrix} gives:
\begin{equation}
	\textbf{b}^\prime=\textbf{P}^{-1} \textbf{b} \\
	=
	\begin{pmatrix} -\frac{7ka^2}{8\zeta_0x} \\
	\frac{kx}{6\zeta_0}(3+3\sqrt{3}) - \frac{ka}{8\zeta_0}(9+5\sqrt{3}) + \frac{7ka^2}{16\zeta_0 x}(1-\sqrt{3}) \\
	\frac{kx}{6\zeta_0}(3-3\sqrt{3}) - \frac{ka}{8\zeta_0}(9-5\sqrt{3}) + \frac{7ka^2}{16\zeta_0 x}(1+\sqrt{3})
	\end{pmatrix}
\end{equation}

The swimming velocity is the first entry of transformed constant vector: $v_{swim} = b^\prime_1 = -\frac{7 k a^2}{8 \zeta_0 x}$. Obviously if we set the initial position of $r_1^\prime$ as $x$, its final position will be the stroke size $x+L$ at transit time T. Thus the dominant path for the center of mass is:
\begin{equation}
	\label{eq:domi1}
	r_1^\prime(t) = -\frac{7ka^2}{8\zeta_0} t + x = - \frac{L}{T} t+ x
\end{equation} 
Equation~\eqref{eq:domi1} also reveals that we can estimate $L$ once we have obtained transit time $T$. The dominant paths for $r_2^\prime(t)$ and $r_3^\prime(t)$ follow equation~\eqref{eq:domipath} by substituting in the corresponding eigenvalue and boundary conditions.
          
\section{Transit time for the model swimmer}

Since we are especially interested in the motion along the unstable mode $r_3^\prime(t)$ that gives rise to the transit time of the catalytic reaction, we must find the boundary conditions $r_3^\prime(0)$ and $r_3^\prime(T)$. The exponential escape motion at the transition state is given by a rate $\lambda_3$ Motion along the eigen coordinate is not parallel to the displacement that determines the free energy barrier $\Delta U^\ddagger$. We thus need to calculate the angle $\phi$ between the energetic reaction coordinate and the unstable mode in the euclidean space spanned by $r_1$, $r_2$ and $r_3$. The direction of unstable mode is proportional to the components of the eigenvector $\mathbf{e}_3$ associated with eigenvalue $\lambda_3$ along the reaction coordinate is $\mathbf{e}_{rxn} = \frac{1}{\sqrt{2}} (-1, 1, 0)$. We find the angle is:
\begin{align*}
\cos{\phi} & = \frac{1}{\sqrt{2}} \frac{(-1,1,0) \cdot ( -2-\sqrt{3}+\frac{a}{x}(3+2\sqrt{3}) ,1+\sqrt{3}+\frac{a}{x}(\frac{-9-5\sqrt{3}}{4}) ,1) }{ | ( -2-\sqrt{3}+\frac{a}{x}(3+2\sqrt{3}) ,1+\sqrt{3}+\frac{a}{x}(\frac{-9-5\sqrt{3}}{4}) ,1)| } \\
&\approx \frac{1}{4\sqrt{2}(3+\sqrt{3})} [ 12 + 8\sqrt{3} - ( 21 + 13\sqrt{3} )\frac{a}{x} ]
\end{align*}

Thus the transit time for the swimmer with proper traversal barrier is:
\begin{equation}
	\label{eq:transittime}
	T = \frac{\ln(2\beta V^\ddagger) + \gamma}{\lambda_3} = \frac{\ln(2\beta \Delta U^\ddagger) + \gamma + \ln[\sqrt{3}(1 - \frac{5a}{4x}) /\cos^2{\phi}] }{\lambda_3}
\end{equation}

\section{Calculation of stroke size}

The product of transit time $T$ expressed in equation ~\eqref{eq:transittime} of swimming speed immediately leads us to the approximation for the stroke size $L$ quoted in the main text:

\begin{align}
	L & =  \frac{\ln(2\beta \Delta U^\dagger) + \gamma + \ln[\sqrt{3}(1 - \frac{5a}{4x}) /\cos^2{\phi}] }{\lambda_3} \times \frac{7ka^2}{8\zeta_0 x} \\
	& = \frac{7a^2[\ln{(2 \beta \Delta U^\ddagger)} + \gamma + \ln(\sqrt{3}/\cos^2{\phi})]}{8\sqrt{3}x}
\end{align}

%\label{}

% If in two-column mode, this environment will change to single-column format so that long equations can be displayed. 
%\begin{widetext}
%$$\mbox{put long equation here}$$
%\end{widetext}

% Figures should be put into the text as floats. 
% Use the graphics or graphicx packages (distributed with LaTeX2e).
% Put the label that you will use with \ref{} command in the braces of the \label{} command.
%
% \begin{figure}
% \end{figure}

% \begin{table}
% \end{tabular}
% \end{table}

%\begin{acknowledgments}
% Put your acknowledgments here.
%\end{acknowledgments}

\end{document}